\documentclass{article}
\usepackage{graphicx} 
\usepackage{amssymb}
\usepackage{amsmath}
\usepackage{draftwatermark}
\newtheorem{theorem}{Theorem}
\newtheorem{lemma}{Lemma}
\newtheorem{corollary}{Corollary}
\newtheorem{remark}{Remark}

\SetWatermarkText{Accepted Manuscript}
\SetWatermarkScale{3}
\title{On the fluxes induced\\ by interacting fields in multiple scattering}

\author{Andreas Kalogeropoulos and Nikolaos L. Tsitsas\\
School of Informatics, Aristotle University of Thessaloniki\\
agkaloger@csd.auth.gr, ntsitsas@csd.auth.gr}
\date{January 2025}

\begin{document}

\maketitle






\begin{abstract}
\noindent Multiple scattering problems involving point-source excitation of lossy media are considered. The sound fluxes induced by the interactions between the scattered fields are analyzed. Theorems connecting interaction intensities (active and reactive) with the Lagrangian densities and the scattering cross sections are established. Physical bounds regarding cross sections ratios are derived, highlighting the contribution of the interaction-induced intensities to the overall intensity. Explicit results for lossless clusters are given. The case of a cluster composed by point-like scatterers is also analyzed.
\end{abstract}








\section{Introduction}

A fundamental aspect of problems related to multiple scattering, is that the field impinging on one of the cluster's scatterers is the accumulation of both the incident field and the fields scattered by the rest of the cluster. Considering that the only observable quantity in the common vicinity of the cluster is the overall field, the separate contribution of each scatterer of the cluster cannot be observed or studied in a direct manner, particularly in the far-field region. Especially, when the incident field is a spherical wave induced by a point source, the accumulated field is the result of an abundance of interactions between fields participating in the same area all of which are radiated in the far-field. The energy transfer process, i.e., the physical mechanism behind the ``journey'' of the energy flux from the incident source, through the common vicinity of the cluster, to the far-field region, is central to the understanding of scattering phenomena. 
     
Investigation of the complex form of acoustic intensity is not new in acoustics \cite{Jacobsen-1991}. Its utilization combined with physical interpretations of the active and reactive intensities \cite{JASA2, Stanzial} have found applications in the calculation of reverberant sound fields \cite{Nolan} and measurement techniques in acoustic waveguides \cite{JASA}, underwater acoustics \cite{Zhang} and in inverse problems related to source localization \cite{Aramini, Chardon}. Aspects of active and reactive intensities have been recently investigated from different standpoints, including their statistical behavior \cite{Jacobsen}, numerical evaluation in enclosures \cite{Meissner} and their connection to the Lagrangian density and scattering cross section (SCS) \cite{PRSA}. On the other hand, the utilization of techniques related to the Green's scalar and vector identities for modeling and analyzing the behavior of energy quantities in wave scattering has its roots in the seminal works of Twersky \cite{Twersky} and has proven a useful tool for the explanation of the energy transfer process since then \cite{Spyropoulos,Tsitsas-SAPM, SAPM, Marston}. The direct multiple-scattering problem of the acoustic excitation of a cluster consisting of fluid spheres by a point source lying in the exterior of the cluster, was analytically solved in \cite{Wu}. Features of the energy transfer process were studied in \cite{RSL} for the electromagnetic excitation of a dielectric, lossless cluster by an external electric dipole. Furthermore, the non-linear nature of the quantities related to the energy and flux has been investigated for multiple scattering problems in \cite{PRSA, Martin, Martin2} and for problems related to layered scatterers in \cite{QAM1}. 
     
In this work, the source that excites the cluster of homogeneous, lossy, penetrable scatterers is enclosed in the interior of a host medium which lies at close proximity to the rest of the scatterers, so that it can be considered a part of the cluster. The interactions between the various elicited fields are accumulating into an overall external field that radiates energy in the far-field region. In multiple scattering problems, only the overall scattered quantities are observable \cite{Martin}. The separate contributions from each scatterer to the overall field and, subsequently, their fluxes, need to be estimated by different means. Here, in an attempt to address this issue, the overall SCS is decomposed into the single-scatterer SCSs, accounting for the flux induced by each single scatterer, when considered unaffected from the interactions by the rest of the cluster, and the cluster-interaction SCS, that accumulates the flux induced by the interactions between different scatterers. The level of participation of the SCSs is analyzed by means of physical bounds related to the ratios of these SCSs to the overall one. These bounds demonstrate the significance of specific single-scatterer SCS and their relation to the number of the cluster's scatterers. The case of a cluster composed by point-like scatterers is explicitly analyzed. Additionally, the corresponding acoustic intensities that quantify the flux in the vicinity of the cluster are also studied and their behavior is studied by means of optical-like theorems and calculation formulas. Aspects of the role of the real part of the acoustic intensity (active intensity) and the imaginary part of the acoustic intensity (reactive intensity) are investigated.

The rest of the paper is structured as follows: In Section 2, the problem under consideration is postulated, the fundamental energy and flux quantities are established and the decomposition of the overall field into single-scatterer fields is presented. In Section 3, theorems connecting the acoustic intensities in the vicinity of the cluster and the flux radiated in the far-field are stated and proved with an emphasis in the behavior of the quantities related to the interaction between participating fields. The contribution of the single-scatterer quantities is studied under the scope of cross-section ratios and their behavior for limiting cases is also examined. In Section 4, by employing a technique stemming from Foldy's approximate method, the behavior of the energy and flux quantities is investigated with a focus in the cluster's response when they are considered point-like with respect to the host medium of the source. Finally, Section 5 summarizes the findings of this work.

\section{Problem formulation}
\subsection{Cluster geometry and overall fields}
\begin{figure}[!htb]
\centering
\includegraphics[width=0.65\textwidth]{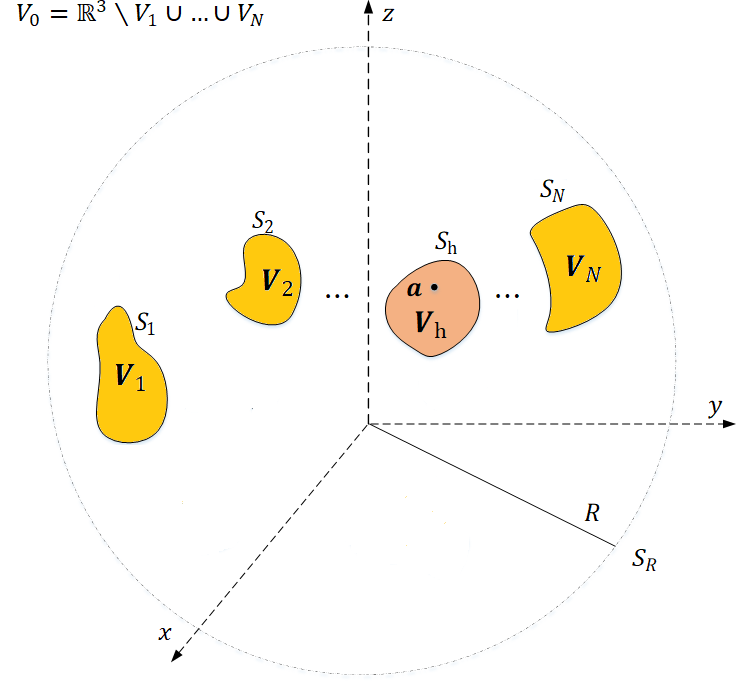}
\caption{A cluster of penetrable scatterers is excited by a point source enclosed inside a penetrable, host medium $V_{h}$.}
\label{fig:mult1}
\end{figure}

Let $V=\cup_{n=1}^NV_n$, for $n=1,\ldots,N$, be a cluster of $N$ homogeneous, penetrable, lossy scatterers with $V_j\cap V_n=\emptyset$ for $j\ne n$. 
The cluster is surrounded by a homogeneous, isotropic and lossless medium $V_0=\mathbb{R}^3\setminus \overline{V}$ characterized by (real) wavenumber $k_0$, mass density $\rho_0$, and mean compressibility $\gamma_0$. 
The penetrable scatterers are characterized by (complex) wavenumbers $k_n$, mass densities $\rho_n$ and mean compressibilities $\gamma_n$, with $n=1,\ldots,N$.
The wavenumbers in $V_0$ and $V_n$ are, respectively, given by
\begin{align}
    \label{wavenumber_0}
    &k_0=\omega\sqrt{\gamma_0\rho_0}, \\
    &k_n=\omega\sqrt{\gamma_n\beta_n},
\label{wavenumber_n}
\end{align}
with 
$$\beta_n=\rho_n\frac{1+\mathrm{i}\nu_n}{1+\nu_n^2},\quad \nu_n=\omega\gamma_n\delta_n$$
and $\delta_n$ the compressional viscosity representing the losses in $V_n$. The choice of the branch of the square root in (\ref{wavenumber_n}) is such that $\mathrm{Im}[k_n]\geq0$ \cite{LFS}.
The cluster is excited by an incident spherical acoustic field due to a point source enclosed in a host-medium $V_h$ with $h\in\{1,\ldots,N\}$; see Fig.~\ref{fig:mult1}. Assuming and suppressing an $\exp(-\mathrm{i}\omega t)$ time dependence, the primary acoustic field has the form
\begin{equation}\label{eq:prim}
    u_h^{\mathrm{pr}}(\mathbf{r})=A\frac{\exp{\left(\mathrm{i}k_h\lvert \mathbf{r}-\mathbf{a}\rvert \right)}}{\lvert \mathbf{r}-\mathbf{a}\rvert }, \quad \mathbf{r}\ne\mathbf{a}.
\end{equation}
%

%

The presence of the cluster perturbs the primary field resulting in the accumulation of an \emph{overall scattered field} $u^{\mathrm{0}}$ incorporating all the fields scattered by the cluster\footnote{assuming that none of the cluster's scatterers lies in the far-field region of the elicited scattered waves.}. On the other hand, in the interior of the penetrable scatterers, the \emph{internal fields} $u^n$, are elicited. For the acoustic field in $V_{h}$, the following decomposition holds
\begin{equation}\label{eq:ps_decomp}
    u^h(\mathbf{r})=u_h^{\mathrm{pr}}(\mathbf{r})+u_h^{\mathrm{sc}}(\mathbf{r}), \quad \mathbf{r}\in V_{h}\setminus\{\mathbf{a}\}.
\end{equation}
The acoustic fields $u^0$, $u^n$ are connected with their respective velocity potentials  $\phi^0$, $\phi^n$ by the relations 
\begin{align}\label{const_0}
    &u^0(\mathbf{r})=\mathrm{i}\omega\rho_0\phi^0(\mathbf{r}), \quad \mathbf{r}\in V_0\\
    \label{const_n}&u^n(\mathbf{r})=\mathrm{i}\omega\beta_n\phi^n(\mathbf{r}), \quad \mathbf{r}\in V_n
\end{align}
%
with Eq.~(\ref{const_n}) holding as well for the respective primary field and potential for $\mathbf{r}\in V_h\setminus\{\mathbf{a}\}$. 
The above fields and velocity potentials satisfy the scalar Helmholtz equations
\begin{equation}\label{Helmholtz}
    (\nabla^2+k_n^2)\begin{Bmatrix}u^n(\mathbf{r})\\\phi^n(\mathbf{r})\end{Bmatrix}=0.
\end{equation}

On the other hand, the transmission boundary conditions for the penetrable scatterers of the cluster read
\begin{align}
\label{transmission}
    u^0(\mathbf{r})&=u^n(\mathbf{r}), \quad \mathbf{r}\in S_n\\
    \varrho_n\frac{\partial u^0(\mathbf{r})}{\partial n}&=\frac{\partial u^n(\mathbf{r})}{\partial n}, \quad \mathbf{r}\in S_n
    \label{BC_pen}
\end{align}
with $n=1,\ldots,N$ and $$\varrho_n=\frac{\beta_n}{\rho_0},$$ a dimensionless constant reducing to the relative mass density in case of a lossless scatterer $V_n$; see (1.26) of \cite{Martin_MS}, while $S_n$ are the boundaries of the cluster's scatterers. 
The Sommerfeld radiation condition reads
\begin{equation}\label{eq:Sommerfeld}
    \lim_{r\rightarrow\infty}r\left(\frac{\partial u^0(\mathbf{r})}{\partial r}-\mathrm{i}k_0u^0(\mathbf{r}) \right)=0.
\end{equation}
Since $V_0$ is unbounded, all solutions of (\ref{Helmholtz}) satisfying (\ref{eq:Sommerfeld}) take the asymptotic form \cite{Colton-Kress-IEM}
\begin{equation}\label{far-field}
    u^0(\mathbf{r})=g^0(\hat{\mathbf{r}})h_0(k_0r)+\mathcal{O}(r^{-2}), \quad r\rightarrow\infty,
\end{equation}
with $\hat{\mathbf{r}}=\mathbf{r}/r$, $h_0$ the $0$-th order spherical Hankel function of the first kind and $g^0(\hat{\mathbf{r}})$ the \emph{far-field pattern}. The $L^2$-norm of the far-field pattern is the \emph{scattering cross section} (SCS), i.e.,
\begin{equation}
\label{scs}
    \sigma=\frac{1}{k_0^2}\int_{S^2}\lvert g^0(\hat{\mathbf{r}})\rvert ^2\mathrm{d}s(\hat{\mathbf{r}}),
\end{equation}
with $S^2$ the unit sphere of $\mathbb{R}^3$.


\subsection{Sound fluxes and energies}
The overall scattered field $u^{0}$ and the internal fields $u^{n}$ are decomposed as
\begin{equation}\label{sum_decomp}
    u^{n}(\mathbf{r})=\sum_{j=1}^{N}u_j^n(\mathbf{r}), \quad \mathbf{r}\in V_n,
\end{equation}
for $n=0,1,\ldots,N$ and $u_j^n$ denoting the \emph{single-scatterer fields}, that quantify \emph{all} contributions from the scatterer $V_j$ of the cluster to the overall field in $V_n$. These fields satisfy the scalar Helmholtz equations in the respective domains as well as the transmission boundary conditions (\ref{transmission})-(\ref{BC_pen}).

In the same spirit, $u_j^\mathrm{0}$ satisfy the Sommerfeld radiation condition (\ref{eq:Sommerfeld}) and take the asymptotic form \cite{Colton-Kress-IEM}
\begin{equation}\label{far-field}
    u_j^0(\mathbf{r})=g_j^0(\hat{\mathbf{r}})h_0(k_0r)+\mathcal{O}(r^{-2}), \quad r\rightarrow\infty,
\end{equation}
with $g_j^0(\hat{\mathbf{r}})$ the far-field pattern due to the $j$-th scatterer. 

The single-scatterer SCSs are given by
\begin{align}\label{iscs}
    &\sigma_j=\frac{1}{k_0^2}\int_{S^2}\lvert g_j^0(\hat{\mathbf{r}})\rvert ^2\mathrm{d}s(\hat{\mathbf{r}}),
\end{align}
and hence the SCS $\sigma$ is written as
\begin{equation}
\label{decomp}
    \sigma=\sigma_\mathrm{c}+\sum_{j=1}^{N}\sigma_j,
\end{equation}
with $\sigma_\mathrm{c}$ the \emph{cluster interaction cross section} given by
\begin{equation}
\label{CICS}
    \sigma_\mathrm{c}=\frac{1}{k_0^2}\sum_{\substack{k,m\\k\ne m}}\int_{S^2} g_k(\hat{\mathbf{r}})\overline{g_m}(\hat{\mathbf{r}})\mathrm{d}s(\hat{\mathbf{r}}),
\end{equation}
where
\begin{equation*}
    \sum_{\substack{k,m\\k\ne m}}a_k\overline{a_m}=\sum_{k=1}^{N}\sum_{m=1}^{N}a_k\overline{a_m}-\sum_{k=1}^{N}\lvert a_k\rvert^2
\end{equation*}
accounting for the flux attributed to the interaction between the single-scatterer fields that remains in quadrature \cite{Stanzial, QAM1}\footnote{This flux can be seen as the reaction between fields elicited by different scatterers of the cluster, a concept introduced in electromagnetics in 1954 \cite{Rumsey} and recently generalized in \cite{Sihvola}.}.  

The scattering and interaction cross sections will be connected in the next section with the flux in the vicinity of the cluster, which is quantified by the overall acoustic intensities $\mathbf{I}^{n}$ in $V_n$. These intensities are defined by
\begin{equation}
    \mathbf{I}^{n}(\mathbf{r})=u^{n}(\mathbf{r})\nabla\overline{\phi^{n}}(\mathbf{r})=\frac{\mathrm{i}}{\omega\overline{\beta_n}}u^{n}(\mathbf{r})\nabla\overline{u^{n}}(\mathbf{r}).
\end{equation}
Taking into account (\ref{sum_decomp}), the last relation takes the form
\begin{equation}\label{intensity_decomp}
    \mathbf{I}^{n}(\mathbf{r})=\mathbf{I}_\mathrm{D}^n(\mathbf{r})+\mathbf{I}_\mathrm{R}^n(\mathbf{r})
\end{equation}
with
\begin{equation}
    \label{direct_intensity}
    \mathbf{I}_\mathrm{D}^n(\mathbf{r})=\frac{\mathrm{i}}{\omega\overline{\beta_n}}\sum_{j=1}^{N}u_j^{n}(\mathbf{r})\nabla\overline{u_j^{n}}(\mathbf{r})
\end{equation}
being the \emph{direct intensity}, i.e., the intensity induced by the single-scatterer fields, and
\begin{equation}
    \label{reaction_intensity}
    \mathbf{I}_{\mathrm{R}}^n(\mathbf{r})=\frac{\mathrm{i}}{\omega\overline{\beta_n}}\sum_{\substack{k,m\\k\ne m}}\left(u_k^{n}(\mathbf{r})\nabla\overline{u_m^{n}}(\mathbf{r})+u_m^{n}(\mathbf{r})\nabla\overline{u_k^{n}}(\mathbf{r})\right)
\end{equation}
the \emph{total interaction intensity} that quantifies the acoustic intensity (flux) induced in $V_n$ by the interaction between the fields elicited by the rest of the cluster. The real (imaginary) part of the acoustic intensities constitutes the \emph{active (reactive) intensity}.
%
%



In the same spirit, the potential and kinetic energies $\mathcal{U}^n$ and $\mathcal{K}^n$ in $V_n$, for $n=0,1,\ldots,N$,  defined by
\begin{align}
    &\mathcal{U}^n(\mathbf{r})=\frac{\gamma_n}{2}\lvert u^n(\mathbf{r})\rvert^2,\\
    &\mathcal{K}^n(\mathbf{r})=\frac{\rho_n}{2}\lvert\nabla\phi^n(\mathbf{r})\rvert^2=\frac{\rho_n}{2\omega^2\lvert \beta_n\rvert^2}\lvert\nabla u^n(\mathbf{r})\rvert^2
\end{align}
have similar decompositions 
\begin{equation}
    \mathcal{U}^n(\mathbf{r})=\mathcal{U}^n_{\mathrm{D}}(\mathbf{r})+\mathcal{U}^n_{\mathrm{R}}(\mathbf{r}), \qquad \mathcal{K}^n(\mathbf{r})=\mathcal{K}^n_{\mathrm{D}}(\mathbf{r})+\mathcal{K}^n_{\mathrm{R}}(\mathbf{r}),
\end{equation}
with the \emph{direct potential energy} and the \emph{direct kinetic energy} given by
\begin{align}\label{potential_direct}
    &\mathcal{U}^n_{\mathrm{D}}(\mathbf{r})=\frac{\gamma_n}{2}\sum_{j=1}^{N}\lvert u_j^n(\mathbf{r})\rvert^2\\
    &\mathcal{K}^n_{\mathrm{D}}(\mathbf{r})=\frac{\rho_n}{2\omega^2\lvert \beta_n\rvert^2}\sum_{j=1}^{N}\lvert\nabla u_j^n(\mathbf{r})\rvert^2,
    \label{kinetic_direct}
\end{align}
while the \emph{interaction potential energy} and the \emph{interaction kinetic energy} are defined as
\begin{align}\label{potential_interaction}
    &\mathcal{U}^n_{\mathrm{R}}(\mathbf{r})=\frac{\gamma_n}{2}\sum_{\substack{k,m\\k\ne m}}\left(u_k^{n}(\mathbf{r})\overline{u_m^{n}}(\mathbf{r})+u_m^{n}(\mathbf{r})\overline{u_k^{n}}(\mathbf{r})\right),\\
    &\mathcal{K}^n_{\mathrm{R}}(\mathbf{r})=\frac{\rho_n}{2\omega^2\lvert \beta_n\rvert^2}\sum_{\substack{k,m\\k\ne m}}\left(\nabla u_k^{n}(\mathbf{r})\nabla\overline{u_m^{n}}(\mathbf{r})+\nabla u_m^{n}(\mathbf{r})\nabla\overline{u_k^{n}}(\mathbf{r})\right).
    \label{kinetic_interaction}
\end{align}

The Lagrangian density $\mathcal{L}^n$ in $V_n$ is defined by
\begin{equation}
    \mathcal{L}^n(\mathbf{r})=\mathcal{K}^n(\mathbf{r})-\mathcal{U}^n(\mathbf{r}),
\end{equation}
and has similar decompositions.

\section{Interaction Fluxes and Cross Sections}
We investigate properties of the interaction intensities in an attempt to delve into the energy transfer process. First, we present results for a lossy cluster, and then we give corresponding reductions for a lossless cluster. It is imperative to note that interaction quantities as well as single-scatterer quantities, are unobservable; only the corresponding overall quantities are observable. Here, they are used as means to a better understanding of the energy transfer process and, in particular, of the underlying mechanism that leads to the accumulation of the observable energy and flux quantities.

\subsection{Lossy cluster}
The first result concerns the connection of the cluster-interaction cross section and the reaction intensity as given in the following theorem.
\begin{theorem}
    \label{Theorem0}
    The cluster-interaction cross section $\sigma_\mathrm{c}$ and the reaction intensity $\mathbf{I}_{\mathrm{R}}^0(\mathbf{r})$ in $V_0$ are connected by the relation
\begin{equation}
\label{Theorem1_0}
\frac{\sigma_\mathrm{c}}{\zeta_0}=\lim_{R\rightarrow\infty}\int_{S_R}\hat{\mathbf{r}}\cdot\mathbf{I}_{\mathrm{R}}^0(\mathbf{r})\mathrm{d}s(\hat{\mathbf{r}})
\end{equation}
with $S_R$ a sphere of radius $R$ surrounding the cluster and $\zeta_0=\sqrt{\rho_0/\gamma_0}$.
\end{theorem}
\textbf{Proof.}
By means of (\ref{reaction_intensity}) for the lossless exterior $V_0$, we have
\begin{equation}
\label{Theorem1_1}
        \hat{\mathbf{r}}\cdot\mathbf{I}_{\mathrm{R}}^0(\mathbf{r})=\frac{\mathrm{i}}{\omega\rho_0}\sum_{\substack{k,m\\k\ne m}}\left(u_k^{0}(\mathbf{r})\frac{\partial}{\partial r}\overline{u_m^{0}}(\mathbf{r})+u_m^{0}(\mathbf{r})\frac{\partial}{\partial r}\overline{u_k^{0}}(\mathbf{r})\right).
\end{equation}
%
Letting $r\rightarrow\infty$ and taking into account the Sommerfeld radiation condition (\ref{eq:Sommerfeld}) as well as the asymptotic form (\ref{far-field}), we obtain
\begin{equation}
    \label{Theorem1_3}
    \hat{\mathbf{r}}\cdot\mathbf{I}_{\mathrm{R}}^0(\mathbf{r})=\frac{2}{\zeta_0}\mathrm{Re}\left[\sum_{\substack{k,m\\k\ne m}}\lvert h_0(k_0r)\rvert^2 g_k^{0}(\hat{\mathbf{r}})\overline{g_m^{0}}(\hat{\mathbf{r}})\right], \quad r\rightarrow\infty,
\end{equation}
uniformly over all directions. Integrating over $S_R$, for $R\rightarrow\infty$, we get
\begin{align}
\nonumber
\lim_{R\rightarrow\infty}\int_{S_R}\hat{\mathbf{r}}\cdot\mathbf{I}_{\mathrm{R}}^0(\mathbf{r})\mathrm{d}s(\hat{\mathbf{r}})=\lim_{R\rightarrow\infty}\int_{S_R}\frac{2}{\zeta_0}\mathrm{Re}\left[\sum_{\substack{k,m\\k\ne m}}\frac{1}{k_0^2R^2} g_k^{0}(\hat{\mathbf{r}})\overline{g_m^{0}}(\hat{\mathbf{r}})\right]\mathrm{d}s(\hat{\mathbf{r}})=\\\frac{2}{\zeta_0 k_0^2}\int_{S^2}\mathrm{Re}\left[\sum_{\substack{k,m\\k\ne m}}g_k^{0}(\hat{\mathbf{r}})\overline{g_m^{0}}(\hat{\mathbf{r}})\right]\mathrm{d}s(\hat{\mathbf{r}}),
\label{Theorem1_4}
\end{align}
which, in conjunction with (\ref{CICS}), yields (\ref{Theorem1_0}). 
\begin{flushright}$\Box$\end{flushright}

Relation (\ref{Theorem1_0}) implies that the interaction intensity, i.e., the acoustic flux generated in $V_0$ by the interaction between the fields of the cluster, is radiated in the far-field as the cluster-interaction cross section, i.e., a portion of the overall SCS.

Now, we give certain theorems that illuminate the contribution of the interaction fields and their relation with the interaction scattering cross-section. 
\begin{theorem}\label{Theorem1}
The cluster-interaction cross section is connected to the active interaction intensity through the host medium and the active interaction kinetic energy densities, by the relation
\begin{equation}
\label{Theorem2_01}
\frac{\sigma_{\mathrm{c}}}{\zeta_0}=\mathrm{Re}\left[\int_{S_h}\hat{\mathbf{n}}\cdot\mathbf{I}_{\mathrm{R}}^{h}(\mathbf{r})\mathrm{d}s(\mathbf{r})\right]-2\omega\sum_{\substack{n=1\\n\ne h}}^{N}\int_{V_n}\mathrm{Im}\left[\frac{\beta_n}{\rho_n}-1\right]\mathcal{K}_{\mathrm{R}}^n(\mathbf{r})\mathrm{d}v(\mathbf{r}),
\end{equation}
while the interaction Lagrangian densities are connected to the reaction intensity and the reactive interaction kinetic energy densities as follows:
\begin{align}
\nonumber
\sum_{\substack{n=0\\n\ne h}}^{N}\int_{V_n}\mathcal{L}_{\mathrm{R}}^n(\mathbf{r})\mathrm{d}v(\mathbf{r})=-\frac{1}{2\omega}\mathrm{Im}\left[\int_{S_h}\hat{\mathbf{n}}\cdot\mathbf{I}_{\mathrm{R}}^{h}(\mathbf{r})\mathrm{d}s(\mathbf{r})\right]-\\\sum_{\substack{n=1\\n\ne h}}^{N}\int_{V_n}\mathrm{Re}\left[\frac{\beta_n}{\rho_n}-1\right]\mathcal{K}_{\mathrm{R}}^n(\mathbf{r})\mathrm{d}v(\mathbf{r}).
\label{Theorem2_02}
\end{align}
\end{theorem}
\textbf{Proof.}
Let $S_R$ be a sphere of radius $R$, centered at the origin, surrounding the scatterers of the cluster and $\Omega_R$ the region bound by $S_R$ and the cluster's boundary $S=\cup_{n=1}^{N}S_n$. Applying the divergence theorem for any of the single-scatterer intensities $\mathbf{I}_j^{0}$ in $\Omega_R$, we obtain
\begin{equation}
\label{Theorem2_1}    \int_{\Omega_R}\nabla\cdot\mathbf{I}_j^0(\mathbf{r})\mathrm{d}v(\mathbf{r})=\int_{S_R}\hat{\mathbf{n}}\cdot\mathbf{I}_j^0(\mathbf{r})\mathrm{d}s(\mathbf{r})-\sum_{n=1}^{N}\int_{S_n}\hat{\mathbf{n}}\cdot\mathbf{I}_j^0(\mathbf{r})\mathrm{d}s(\mathbf{r}).
\end{equation}
For the integrand in the left-hand side of (\ref{Theorem2_1}) it holds
\begin{equation}
\nabla\cdot\mathbf{I}_j^0(\mathbf{r})=\frac{\mathrm{i}}{\omega\rho_0}\nabla\cdot\left(u_j^0(\mathbf{r})\nabla\overline{u_j^0}(\mathbf{r})\right)
\end{equation}
and taking into account that $u_j^0$ are solutions of the Helmholtz equation in $\Omega_R$ we obtain
\begin{equation}\label{Theorem2_2}
    \nabla\cdot\mathbf{I}_j^0(\mathbf{r})=2\mathrm{i}\omega\left(\frac{1}{2\omega^2\rho_0}\left|\nabla u_j^0(\mathbf{r})\right|^2-\frac{\gamma_0}{2}\left|u_j^0(\mathbf{r})\right|^2\right)=2\mathrm{i}\omega\mathcal{L}_j^0(\mathbf{r}),
\end{equation}
with $\mathcal{L}_j^0(\mathbf{r})$, the Lagrangian density in $V_0$ attributed to the $j$-th scatterer of the cluster.

On the other hand, letting $r\rightarrow\infty$ and taking into account (\ref{eq:Sommerfeld}), we arrive at the following asymptotic relation

\begin{equation}\label{Theorem2_3}
    \hat{\mathbf{r}}\cdot\mathbf{I}_j^0(\mathbf{r})=\frac{\lvert g_j^0(\hat{\mathbf{r}})\rvert^2}{\zeta_0}\frac{1}{k_0^2r^2}, \qquad r\rightarrow\infty.
\end{equation}
Utilizing Green's first identity for each integral of the second term in the right-hand side of (\ref{Theorem2_1}) we find
\begin{equation}
    \label{Theorem2_4}
    \int_{S_n}\hat{\mathbf{n}}\cdot\mathbf{I}_j^0(\mathbf{r})\mathrm{d}s(\mathbf{r})=\int_{V_n}\nabla\cdot\mathbf{I}_j^n(\mathbf{r})\mathrm{d}v(\mathbf{r}).
\end{equation}
Taking into account (\ref{const_n}), the integrand in the right-hand side of (\ref{Theorem2_4}) becomes
\begin{equation}
    \label{Theorem2_5}
    \nabla\cdot\mathbf{I}_j^n(\mathbf{r})=2\mathrm{i}\omega\left(\mathcal{L}_j^n(\mathbf{r})+\left(\frac{\beta_n}{\rho_n}-1\right)\mathcal{K}_j^n(\mathbf{r})\right)
\end{equation}
for $n=1,\ldots,N$, and $n\ne h$, while in the host medium (i.e, for $n=h$) it holds
\begin{equation}
    \label{Theorem2_6}
    \int_{S_h}\hat{\mathbf{n}}\cdot\mathbf{I}_j^0(\mathbf{r})\mathrm{d}s(\mathbf{r})=\int_{S_h}\hat{\mathbf{n}}\cdot\mathbf{I}_j^h(\mathbf{r})\mathrm{d}s(\mathbf{r}).
\end{equation}   

Substituting (\ref{Theorem2_2})-(\ref{Theorem2_6}) in (\ref{Theorem2_1}), utilizing the definition (\ref{iscs}) of the single-scatterer cross section and considering that for $R\rightarrow\infty$ region $\Omega_R$ coincides with $V_0$, we arrive at
%
%
\begin{align}
    \nonumber
    2\mathrm{i}\omega\sum_{\substack{n=0\\n\ne h}}^{N}\int_{V_n}\mathcal{L}_j^n(\mathbf{r})\mathrm{d}v(\mathbf{r})=\frac{\sigma_j}{\zeta_0}-\int_{S_h}\hat{\mathbf{n}}\cdot\mathbf{I}_j^h(\mathbf{r})\mathrm{d}s(\mathbf{r})-\\2\mathrm{i}\omega\sum_{\substack{n=1\\n\ne h}}^{N}\int_{V_n}\left(\frac{\beta_n}{\rho_n}-1\right)\mathcal{K}_j^n(\mathbf{r})\mathrm{d}v(\mathbf{r}).
    \label{Theorem2_8}
\end{align}
Following a similar procedure for the overall acoustic intensity, we obtain
\begin{align} 
\nonumber
2\mathrm{i}\omega\sum_{\substack{n=0\\n\ne h}}^{N}\int_{V_n}\mathcal{L}^n(\mathbf{r})\mathrm{d}v(\mathbf{r})=\frac{\sigma}{\zeta_0}-\int_{S_h}\hat{\mathbf{n}}\cdot\mathbf{I}^{h}(\mathbf{r})\mathrm{d}s(\mathbf{r})-\\2\mathrm{i}\omega\sum_{\substack{n=1\\n\ne h}}^{N}\int_{V_n}\left(\frac{\beta_n}{\rho_n}-1\right)\mathcal{K}^n(\mathbf{r})\mathrm{d}v(\mathbf{r}).
\label{Theorem2_9}
\end{align}
Relation (\ref{Theorem2_8}) holds for all the single-scatterer fields (including the host-medium) and, thus, equation (\ref{Theorem2_9}) in combination with (\ref{CICS}) and (\ref{intensity_decomp}) lead to
\begin{align} \nonumber
2\mathrm{i}\omega\sum_{\substack{n=0\\n\ne h}}^{N}\int_{V_n}\mathcal{L}_{\mathrm{R}}^n(\mathbf{r})\mathrm{d}v(\mathbf{r})=\frac{\sigma_{\mathrm{c}}}{\zeta_0}-\int_{S_h}\hat{\mathbf{n}}\cdot\mathbf{I}_{\mathrm{R}}^{h}(\mathbf{r})\mathrm{d}s(\mathbf{r})-\\2\mathrm{i}\omega\sum_{\substack{n=1\\n\ne h}}^{N}\int_{V_n}\left(\frac{\beta_n}{\rho_n}-1\right)\mathcal{K}_{\mathrm{R}}^n(\mathbf{r})\mathrm{d}v(\mathbf{r}).\label{Theorem2_10}
\end{align}
Taking the real parts of (\ref{Theorem2_10}) leads to (\ref{Theorem2_01}), while taking the imaginary parts leads to (\ref{Theorem2_02}).
 \begin{flushright}$\Box$\end{flushright}

By considering the limiting case of $\omega\rightarrow0$, we obtain the following corollaries
\begin{corollary}\label{corol2}
    For $\omega\rightarrow0$, (\ref{Theorem2_01}) and (\ref{Theorem2_02}) take the following asymptotic forms, respectively
\begin{align}
   \label{Corollary2_01}
       &\frac{\sigma_\mathrm{c}}{\zeta_0}\rightarrow\mathrm{Re}\left[\int_{S_h}\hat{\mathbf{n}}\cdot\mathbf{I}_{\mathrm{R}}^{h}(\mathbf{r})\mathrm{d}s(\mathbf{r})\right], \quad \omega\rightarrow0\\
       \label{Corollary2_02}
        &\mathrm{Im}\left[\int_{S_h}\hat{\mathbf{n}}\cdot\mathbf{I}_{\mathrm{R}}^{h}(\mathbf{r})\mathrm{d}s(\mathbf{r})\right]\rightarrow0, \quad \omega\rightarrow0.
        \end{align}
\end{corollary}

Equation (\ref{Theorem2_01}) connects the cluster-interaction cross section, accounting for the flux radiated in the far-field, to the active interaction intensity out of the host-medium and the interaction losses in the rest of the cluster. Another interesting observation about the cluster-interaction cross section, is that it can be negative, contrary to the single-scatterer cross sections and the overall cross section. This has been observed for layered, spherical scatterers in electromagnetics \cite{SAPM,OJAP} and in acoustics \cite{QAM1} for the total and indirect interaction cross sections, two quantities that quantify the various interactions between fields induced in different layers and between all the fields impinging on the layered scatterer. Equation (\ref{Theorem2_01}) implies that the cluster-interaction cross section is negative ($\sigma_\mathrm{c}<0$) if and only if the interaction intensity through the surface of the host medium is less than the sum of the interaction kinetic energies in the rest of the cluster, i.e., if the following condition holds
\begin{equation}
    \label{Theorem2_05}
    \mathrm{Re}\left[\int_{S_h}\hat{\mathbf{n}}\cdot\mathbf{I}_{\mathrm{R}}^{h}(\mathbf{r})\mathrm{d}s(\mathbf{r})\right]<2\omega\sum_{\substack{n=1\\n\ne h}}^{N}\int_{V_n}\mathrm{Im}\left[\frac{\beta_n}{\rho_n}-1\right]\mathcal{K}_{\mathrm{R}}^n(\mathbf{r})\mathrm{d}v(\mathbf{r}).
\end{equation}
The right hand side of the above inequality vanishes for lossless scatterers.

Now, we provide a formula that connects the active reaction intensity out of the host medium with the point source's strength and the specific admittance $\zeta_h=\left(\mathrm{Re}\left[{\frac{k_h}{\beta_h}}\right]\right)^{-1}$ of the host medium.
%
\begin{theorem}
    For the active reaction intensity $\mathbf{I}_{\mathrm{R}}^h$ out of the host medium, it holds
\begin{equation}
\label{Theorem6_0}
\int_{S_h}\mathrm{Re}\left[\hat{\mathbf{r}}\cdot\mathbf{I}_{\mathrm{R}}^h(\mathbf{r})\right]\mathrm{d}s(\mathbf{r})=\frac{4\pi\lvert A\rvert^2}{\zeta_h}-2\omega\mathrm{Im}\left[\frac{\beta_h}{\rho_h}\right]\int_{V_h}\mathcal{K}_{\mathrm{R}}^h(\mathbf{r})\mathrm{d}v(\mathbf{r}).
\end{equation}
\end{theorem}
\textbf{Proof.}
    Let $\mathbf{I}_j^h$ be any of the single-scatterer intensities, $B(\mathbf{a};\epsilon)$ an open ball of radius $\epsilon$ centered at the position $\mathbf{a}$ of the source with $B(\mathbf{a};\epsilon)\subseteq V_h$ and $\Omega_h=V_h\setminus \overline{B}(\mathbf{a};\epsilon)$. Applying Green's first identity for $\mathbf{I}_j^h$ in $\Omega_h$, yields
\begin{align}
\nonumber
\frac{\mathrm{i}}{\omega\overline{\beta_h}}\int_{\Omega_h}\left(\lvert\nabla u_j^{h}(\mathbf{r})\rvert^2-\overline{k^2}\lvert u_j^{h}(\mathbf{r})\rvert^2\right)\mathrm{d}v(\mathbf{r})=\\\int_{S_h}\hat{\mathbf{r}}\cdot\mathbf{I}_j^h(\mathbf{r})\mathrm{d}s(\mathbf{r})-\int_{\partial B(\mathbf{a};\epsilon)}\hat{\mathbf{r}}\cdot\mathbf{I}_j^h(\mathbf{r})\mathrm{d}s(\mathbf{r})\label{Theorem6_1}
\end{align}
Taking the real parts of (\ref{Theorem6_1}), and after some manipulation, we obtain
\begin{align}
\nonumber
\int_{S_h}\mathrm{Re}\left[\hat{\mathbf{r}}\cdot\mathbf{I}_j^h(\mathbf{r})\right]\mathrm{d}s(\mathbf{r})=\int_{\partial B(\mathbf{a};\epsilon)}\mathrm{Re}\left[\hat{\mathbf{r}}\cdot\mathbf{I}_j^h(\mathbf{r})\right]\mathrm{d}s(\mathbf{r})-\\2\omega\mathrm{Im}\left[\frac{\beta_h}{\rho_h}\right]\int_{\Omega_h}\mathcal{K}_j^h(\mathbf{r})\mathrm{d}v(\mathbf{r}).\label{Theorem6_3}
\end{align}

In the host medium $V_h$, the following decompositions are valid
\begin{align}
\label{host_decomp_single}
\mathbf{I}_j^h(\mathbf{r})&=\mathbf{I}_j^{\mathrm{sec}}(\mathbf{r})+\mathbf{I}_j^{\mathrm{ext}}(\mathbf{r})\\
\label{host_decomp_overall}
\mathbf{I}^h(\mathbf{r})&=\mathbf{I}^{\mathrm{pr}}(\mathbf{r})+\mathbf{I}^{\mathrm{sec}}(\mathbf{r})+\mathbf{I}^{\mathrm{ext}}(\mathbf{r}),
\end{align}
where
\begin{align*}
&\mathbf{I}_j^{\mathrm{sec}}(\mathbf{r})=\frac{\mathrm{i}}{\omega\overline{\beta_h}}u_j^{h}(\mathbf{r})\nabla\overline{u_j^{h}}(\mathbf{r}), \\&\mathbf{I}_j^{\mathrm{ext}}(\mathbf{r})=\frac{\mathrm{i}}{\omega\overline{\beta_h}}\left(u_j^{h}(\mathbf{r})\nabla\overline{u^{\mathrm{pr}}}(\mathbf{r})+u^{\mathrm{pr}}(\mathbf{r})\nabla\overline{u_j^{h}}(\mathbf{r})\right),\\
&\mathbf{I}^{\mathrm{pr}}(\mathbf{r})=\frac{\mathrm{i}}{\omega\overline{\beta_h}}u^{\mathrm{pr}}(\mathbf{r})\nabla\overline{u^{\mathrm{pr}}}(\mathbf{r}), \quad \mathbf{I}^{\mathrm{sec}}(\mathbf{r})=\sum_{j=1}^{N}\mathbf{I}_j^{\mathrm{sec}}(\mathbf{r}), \quad \mathbf{I}^{\mathrm{ext}}(\mathbf{r})=\sum_{j=1}^{N}\mathbf{I}_j^{\mathrm{ext}}(\mathbf{r}).
\end{align*}
For $\mathbf{I}_j^{\mathrm{ext}}$, we have
\begin{align}
\nonumber
\hat{\mathbf{r}}\cdot\mathbf{I}_j^{\mathrm{ext}}(\mathbf{r})=\frac{\mathrm{i}}{\omega\overline{\beta_h}}\Bigg(A\frac{\exp{(\mathrm{i}k_h\lvert\mathbf{r}-\mathbf{a}\rvert)}}{\lvert\mathbf{r}-\mathbf{a}\rvert}\frac{\partial \overline{u_j^{h}}(\mathbf{r})}{\partial r}-\\\overline{A}u_j^{h}(\mathbf{r})\exp{(-\mathrm{i}\overline{k_h}\lvert\mathbf{r}-\mathbf{a}\rvert)}\frac{\mathrm{i}\overline{k_h}\lvert\mathbf{r}-\mathbf{a}\rvert+1}{\lvert\mathbf{r}-\mathbf{a}\rvert^2}\Bigg).
\label{Theorem6_4}
\end{align}
Applying the mean value theorem in $B(\mathbf{a};\epsilon)$, and letting $\epsilon\rightarrow0$, we get
\begin{equation}
    \label{Theorem6_5}
    \lim_{\epsilon\rightarrow0}\int_{\partial B(\mathbf{a};\epsilon)}\mathrm{Re}\left[\hat{\mathbf{r}}\cdot\mathbf{I}_j^{\mathrm{ext}}(\mathbf{r})\right]\mathrm{d}s(\hat{\mathbf{r}})=-\frac{4\pi}{\omega}\mathrm{Im}\left[\frac{A}{\beta_h}\overline{u_j^{h}}(\mathbf{a})\right].
\end{equation}
Similarly, for $\mathbf{I}_j^{\mathrm{sc}}$, we obtain
\begin{equation}
\label{Theorem6_6}
\int_{\partial B(\mathbf{a};\epsilon)}\mathrm{Re}\left[\hat{\mathbf{r}}\cdot\mathbf{I}_j^{\mathrm{sc}}(\mathbf{r})\right]\mathrm{d}s(\hat{\mathbf{r}})=0.
\end{equation}
Equations (\ref{Theorem6_5}) and (\ref{Theorem6_6}) in combination with (\ref{Theorem6_3}) and decomposition (\ref{host_decomp_single}) lead to
\begin{align}
\nonumber
\int_{S_h}\mathrm{Re}\left[\hat{\mathbf{r}}\cdot\mathbf{I}_j^h(\mathbf{r})\right]\mathrm{d}s(\mathbf{r})=-\frac{4\pi}{\omega}\mathrm{Im}\left[\frac{A}{\beta_h}\overline{u_j^{h}}(\mathbf{a})\right]-\\2\omega\mathrm{Im}\left[\frac{\beta_h}{\rho_h}\right]\int_{\Omega_h}\mathcal{K}_j^h(\mathbf{r})\mathrm{d}v(\mathbf{r}).\label{Theorem6_7}
\end{align}
For the primary intensity, taking into account (\ref{eq:prim}), it holds
\begin{equation}
\label{Theorem6_8}\mathrm{Re}\left[\hat{\mathbf{r}}\cdot\mathbf{I}^{\mathrm{pr}}(\mathbf{r})\right]=\mathrm{Re}\left[\frac{\mathrm{i}}{\omega\overline{\beta_h}}u^{\mathrm{pr}}(\mathbf{r})\frac{\partial\overline{u^{\mathrm{pr}}}(\mathbf{r})}{\partial r}\right]=\frac{\lvert A\rvert^2}{\lvert\mathbf{r}-\mathbf{a}\rvert^2}\frac{1}{\zeta_h}.
\end{equation}
Thus, applying the mean value theorem in $B(\mathbf{a};\epsilon)$, and letting $\epsilon\rightarrow0$, we arrive at
\begin{equation}
\label{Theorem6_9}
\lim_{\epsilon\rightarrow0}\int_{\partial B(\mathbf{a};\epsilon)}\mathrm{Re}\left[\hat{\mathbf{r}}\cdot\mathbf{I}^{\mathrm{pr}}(\mathbf{r})\right]\mathrm{d}s(\hat{\mathbf{r}})=\frac{4\pi\lvert A\rvert^2}{\zeta_h}.
\end{equation}
Following a similar path for the intensities $\mathbf{I}^{\mathrm{ext}}(\mathbf{r})$ and $\mathbf{I}^{\mathrm{sc}}(\mathbf{r})$, we get
\begin{align}
        \nonumber
        \int_{S_h}\mathrm{Re}\left[\hat{\mathbf{r}}\cdot\mathbf{I}^h(\mathbf{r})\right]\mathrm{d}s(\mathbf{r})=\frac{4\pi\lvert A\rvert^2}{\zeta_h}\left(1-\frac{\zeta_h}{\omega}\mathrm{Im}\left[\frac{A}{\beta_h}\overline{u_h^{\mathrm{sc}}}(\mathbf{a})\right]\right)-\\2\omega\mathrm{Im}\left[\frac{\beta_h}{\rho_h}\right]\int_{\Omega_h}\mathcal{K}^h(\mathbf{r})\mathrm{d}v(\mathbf{r}).\label{Theorem6_10}
    \end{align}
Combining (\ref{Theorem6_9}) and (\ref{Theorem6_10}) with (\ref{sum_decomp}) and (\ref{host_decomp_overall}) and since, for $\epsilon\rightarrow0$ the domain $\Omega_h$ coincides with $V_h$, we find (\ref{Theorem6_0}).
 \begin{flushright}$\Box$\end{flushright}

Utilizing (\ref{Theorem6_0}), we get an alternative version of (\ref{Theorem2_01})
\begin{equation}\label{Theorem2_03}
    \frac{\sigma_{\mathrm{c}}}{\zeta_0}=\frac{4\pi\lvert A\rvert^2}{\zeta_h}-2\omega\sum_{n=1}^{N}\int_{V_n}\mathrm{Im}\left[\frac{\beta_n}{\rho_n}\right]\mathcal{K}_{\mathrm{R}}^n(\mathbf{r})\mathrm{d}v(\mathbf{r})+2\omega\int_{V_h}\mathcal{K}_{\mathrm{R}}^h(\mathbf{r})\mathrm{d}v(\mathbf{r})
\end{equation}
that connects the cluster-interaction SCS with the kinetic energies of the cluster. 

We note that quantity $\frac{4\pi\lvert A\rvert^2}{\zeta_h}$ constitutes the \emph{primary cross section} $\sigma^{\mathrm{pr}}$, i.e., the flux induced by the point source in the case where the entire $\mathbb{R}^3$ is filled by the material of $V_h$. 

\subsection{Reductions to a lossless cluster}
For a cluster comprised by lossless scatterers, the compressional viscosities $\delta_n$ are zero and thus it holds $\beta_n=\overline{\beta_n}=\rho_n$, for $n=0,1,\ldots,N$. The following corollaries stem from Theorems 2 and 3, respectively.
\begin{corollary}\label{corol1}
For a cluster of lossless scatterers, equations (\ref{Theorem2_01}) and (\ref{Theorem2_02}) take the form, respectively
\begin{align}
   \label{Corollary1_01}
       &\frac{\sigma_\mathrm{c}}{\zeta_0}=\mathrm{Re}\left[\int_{S_h}\hat{\mathbf{n}}\cdot\mathbf{I}_{\mathrm{R}}^{h}(\mathbf{r})\mathrm{d}s(\mathbf{r})\right], \\
       \label{Corollary1_02}
        &2\omega\sum_{\substack{n=0\\n\ne h}}^{N}\int_{V_n}\mathcal{L}_{\mathrm{R}}^n(\mathbf{r})\mathrm{d}v(\mathbf{r})=-\mathrm{Im}\left[\int_{S_h}\hat{\mathbf{n}}\cdot\mathbf{I}_{\mathrm{R}}^{h}(\mathbf{r})\mathrm{d}s(\mathbf{r})\right].
        \end{align}
\end{corollary}
\begin{corollary}\label{corol3}
    For a cluster of lossless scatterers the cluster-interaction cross section and the overall SCS are given by
    \begin{align}
    \label{Theorem6_01}
    &\sigma_\mathrm{c}=4\pi\lvert A\rvert^2\frac{\zeta_0}{\zeta_h}.
\\
\label{OSCS}
    &\sigma=4\pi\lvert A\rvert^2\frac{\zeta_0}{\zeta_h}\left(1-\frac{\zeta_h}{\omega}\mathrm{Im}\left[\frac{A}{\beta_h}\overline{u_h^{\mathrm{sc}}}(\mathbf{a})\right]\right)
\end{align}
\end{corollary}
%
%
%
\begin{remark}
    The results in corollaries \ref{corol1} and \ref{corol3} also apply for a cluster comprised by sound-soft and/or sound-hard scatterers--host-medium excluded--since all acoustic intensities vanish at their surfaces.
\end{remark}

\subsection{Cross-section ratios}
The contributions of the cluster interaction SCS and the sum of the single-scatterer SCSs, are quantified by their corresponding ratios to the overall SCS given by
\begin{equation}\label{ratios}
    \mathrm{R}_{n}=\frac{\sigma_n}{\sigma}, \qquad \mathrm{R}_{\mathrm{c}}=\frac{\sigma_\mathrm{c}}{\sigma}.
\end{equation}
The following results, can be used to estimate accurately the amount of scatterers required to obtain a specific cross section value - especially when the cluster is composed by scatterers that individually radiate similar single-scatterer cross sections. 
\begin{theorem}\label{Theorem3}
   The ratio $R_\mathrm{c}$ of the cluster interaction SCS to the overall SCS and the minimum single-scatterer SCS, $\mathrm{R}_{\mathrm{min}}$, satisfy
    \begin{equation}\label{Theorem3_01}
        R_\mathrm{c}\leq\min\left\{\frac{N-1}{N},1-N\mathrm{R}_{\mathrm{min}}\right\}.
    \end{equation}
\end{theorem}
\textbf{Proof.}
For the maximum and minimum single-scatterer SCSs, $\sigma_{\mathrm{min}}$ and $\sigma_{\mathrm{max}}$, it holds
    \begin{equation}
        \label{Theorem3_1}
        N\sigma_{\mathrm{min}}\leq \sum_{n=1}^{N}\sigma_n\leq N\sigma_{\mathrm{max}},
    \end{equation}
which implies that
\begin{equation}
    \label{Theorem3_2}
    1-N\mathrm{R}_{\mathrm{max}}\leq\mathrm{R}_\mathrm{c}\leq1-N\mathrm{R}_{\mathrm{min}}. 
\end{equation}
On the other hand, for the cluster-interaction cross section by means of (\ref{CICS}), we have
\begin{equation}
    \label{Theorem3_3}
    \sigma_\mathrm{c}\leq\frac{2}{k_0^2}\sum_{n=1}^{N-1}\sum_{m=n+1}^{N}\int_{S^2} \lvert g_n(\hat{\mathbf{r}})\overline{g_m}(\hat{\mathbf{r}})\rvert\mathrm{d}s(\hat{\mathbf{r}})
\end{equation}
which by means of H\"older's inequality yields
\begin{align}
\nonumber
    \sigma_\mathrm{c}\leq\frac{2}{k_0^2}\mathrm{Re}\left[\sum_{n=1}^{N-1}\sum_{m=n+1}^{N}\left(\int_{S^2}\lvert g_n(\hat{\mathbf{r}})\rvert^2\mathrm{d}s(\hat{\mathbf{r}})\right)^{1/2}\left(\int_{S^2}\lvert g_m(\hat{\mathbf{r}})\rvert^2\mathrm{d}s(\hat{\mathbf{r}})\right)^{1/2}\right]=\\2\sum_{n=1}^{N-1}\sum_{m=n+1}^{N}\left(\sqrt{\sigma_n}\sqrt{\sigma_m}\right),
    \label{Theorem3_4}
\end{align}
%
and, thus, we get
\begin{equation}
    \label{Theorem3_5}
    \sigma_\mathrm{c}\leq\sum_{n=1}^{N-1}\sum_{m=n+1}^{N}\left(\sigma_n+\sigma_m\right).
\end{equation}
But, it holds that
\begin{equation}
    \label{Theorem3_6}
    \sum_{n=1}^{N-1}\sum_{m=n+1}^{N}\sigma_n=\sum_{n=1}^{N-1}(N-n)\sigma_n, \qquad \sum_{n=1}^{N-1}\sum_{m=n+1}^{N}\sigma_m=\sum_{n=2}^{N}(n-1)\sigma_n.
\end{equation}
Hence, combining equations (\ref{Theorem3_5}) and (\ref{Theorem3_6}), we arrive at
\begin{equation}
    \label{Theorem3_7}
    \sigma_\mathrm{c}\leq(N-1)\sum_{n=1}^{N}\sigma_n,
\end{equation}
which, by means of (\ref{decomp}), yields
\begin{equation}
    \label{Theorem3_8}
    R_\mathrm{c}\leq\frac{N-1}{N}.    
\end{equation}
Equations (\ref{Theorem3_2}) and (\ref{Theorem3_8}) prove the theorem.
%
\begin{flushright}$\Box$\end{flushright}

\begin{corollary}\label{corol4}
The number $N$ of the cluster's scatterers is bounded by
\begin{equation}\label{corollary3_01}
        \frac{1}{\sqrt{\mathrm{R}_{\mathrm{max}}}}\leq N\leq\frac{1}{\sqrt{\mathrm{R}_{\mathrm{min}}}},
    \end{equation}
if and only if
\begin{equation}\label{corol3_00}
    \frac{N-1}{N}\leq1-N\mathrm{R}_{\mathrm{min}},
\end{equation} 
%
%
with $\mathrm{R}_{\mathrm{max}}$ the maximum single-scatterer cross section.
\end{corollary}
\textbf{Proof.}
The upper bound of $N$ is trivial. For the lower bound, we start from the overall SCS, satisfying
\begin{equation}
\label{corollary3_1}
\sigma\leq N\sigma_{\mathrm{max}}+\sigma_\mathrm{c},
\end{equation}
which, by means of (\ref{Theorem3_5}), takes the form
\begin{equation}
\label{corollary3_2}
\sigma\leq N\sigma_{\mathrm{max}}+2\sum_{n=1}^{N-1}\sum_{m=n+1}^{N}\sigma_{\mathrm{max}}=N^2\sigma_{\mathrm{max}},
\end{equation}
%
%
yielding
\begin{equation}
\label{corollary3_4}
\frac{1}{N^2}\leq \mathrm{R}_{\mathrm{max}}.
\end{equation}
%
 \begin{flushright}$\Box$\end{flushright}

Utilizing Corollary \ref{corol4}, we find
\begin{theorem}\label{Theorem4}
    It holds that
    \begin{equation}
    \label{Theorem4_01}
        \mathrm{R}_{\mathrm{max}}=\frac{1}{N^2}\end{equation}
        if and only if
        \begin{equation} \label{Theorem4_02}   
        \mathrm{R}_{n}=\frac{1}{N^2},\quad \mbox{\rm{for all}} \quad n=1,\ldots,N.
    \end{equation}
\end{theorem}
\textbf{Proof.}
If (\ref{Theorem4_01}) holds then each ratio $\mathrm{R}_j$ can be written as 
$$\mathrm{R}_{j}=\frac{a_j}{N^2}$$
with $0\leq a_j\leq1$ and $a_k=1$ for some $k\in\{1,\ldots,N\}$. Then, it follows that
\begin{equation}
\label{Theorem4_1}
\sum_{n=1}^{N}\mathrm{R}_j=\frac{1}{N^2}\sum_{n=1}^{N}a_j\leq\frac{1}{N},
\end{equation} 
which by means of (\ref{decomp}) and (\ref{Theorem3_8}), yields
$$\mathrm{R}_{\mathrm{c}}=\frac{N-1}{N}.$$
The last relation, combined with (\ref{decomp}) and (\ref{Theorem4_1}), leads to 
$$\sum_{j=1}^{N}a_j=N,$$ 
which readily implies that $a_1=\ldots=a_N=1$.
The inverse statement is obvious.
 \begin{flushright}$\Box$\end{flushright}
\begin{remark}
    An implication of the proof of Theorem (\ref{Theorem4}) is that if $\mathrm{R}_{\mathrm{max}}=\frac{1}{N^2}$ then $\mathrm{R}_{\mathrm{c}}=\frac{N-1}{N}$.
\end{remark} 

Another interesting aspect is the behavior of the SCS (and its ratios) as the number $N$ of the scatterers varies. Without loss of generality, we consider that $\sigma_1\leq\sigma_2\leq\ldots\leq\sigma_N$. 
Furthermore, we denote with $\sigma^{N}$ the overall SCS, when the cluster is composed by $N$ scatterers, and with $\sigma_n^{N-1}$ the overall SCS when the $n$-th largest single-scatterer contribution of the cluster, is removed.
\begin{theorem}\label{theorem6}
    For a cluster composed by scatterers with $\sigma_1\leq\ldots\leq\sigma_N$, it holds
    \begin{equation}
        \label{theorem8_00}
        \sigma^{N}-
\sigma_n^{N-1}\leq(2n-1)\sigma_n+2(N-n-1)\sigma_N.
    \end{equation}
\end{theorem}
\textbf{Proof.}
The total contribution of the scatterer with the $n$-th largest SCS is given by the difference $\sigma^{N}-\sigma_n^{N-1}$, satisfying
\begin{align}
\nonumber
\sigma^{N}-\sigma_n^{N-1}=\frac{1}{k_0^2}\int_{S^2}\left\lvert \sum_{j=1}^{N}g_j(\hat{\mathbf{r}})\right\rvert^2\mathrm{d}s(\hat{\mathbf{r}})-
\frac{1}{k_0^2}\int_{S^2}\left\lvert \sum_{\substack{j=1\\j\ne n}}^{N}g_j(\hat{\mathbf{r}})\right\rvert^2\mathrm{d}s(\hat{\mathbf{r}})=\\\frac{1}{k_0^2}\int_{S^2}\lvert g_n(\hat{\mathbf{r}})\rvert^2\mathrm{d}s(\hat{\mathbf{r}})+\frac{2}{k_0^2}\mathrm{Re}\left[\int_{S^2} g_n(\hat{\mathbf{r}})\sum_{\substack{j=1\\j\ne n}}^{N}\overline{g_j}(\hat{\mathbf{r}})\mathrm{d}s(\hat{\mathbf{r}})\right].
\label{theorem8_1}
\end{align}
Using H\"older's inequality and taking into account (\ref{iscs}), we obtain
\begin{equation}\label{theorem8_2}
\frac{2}{k_0^2}\mathrm{Re}\left[\int_{S^2} g_n(\hat{\mathbf{r}})\sum_{\substack{j=1\\j\ne n}}^{N}\overline{g_j}(\hat{\mathbf{r}})\mathrm{d}s(\hat{\mathbf{r}})\right]\leq2\sum_{\substack{j=1\\j\ne n}}^{N}\sqrt{\sigma_n}\sqrt{\sigma_j}.
\end{equation}
Combining (\ref{theorem8_1}) and (\ref{theorem8_2}) yields
\begin{align}
\nonumber
\sigma^{N}-
\sigma_n^{N-1}\leq\sigma_{n}+2\sqrt{\sigma_n}\left(\sum_{j=1}^{n-1}\sqrt{\sigma_n}+\sum_{j=n+1}^{N}\sqrt{\sigma_j}\right)=\\(2n-1)\sigma_n+2(N-n-1)\sigma_N.\label{theorem8_3}
\end{align}
 \begin{flushright}$\Box$\end{flushright}
Relation (\ref{theorem8_00}) for $n=1$, corresponding to the scatterer with the smallest SCS, and for $n=N$, corresponding to the scatterer with the largest SCS, yields respectively,
\begin{align}
        \label{theorem8_01}
        &\sigma^{N}-\sigma_1^{N-1}\leq \sigma_1+2(N-2)\sigma_N\\
        \label{theorem8_02}
        &\sigma^{N}-\sigma_N^{N-1}\leq (2N-3)\sigma_N
\end{align}

An interesting observation stems from a different handling of (\ref{theorem8_3}) and considering $N\sigma_n\leq\sigma_{\mathrm{D}}^{N}$, yielding
\begin{align}
\label{theorem8_4}
\sigma^{N}\leq
\sigma_n^{N-1}+2\sigma_{\mathrm{D}}^{N},
\end{align}
which leads to
\begin{align}
\label{theorem8_5}
R_{\mathrm{c}}^{N}-R_{\mathrm{D}}^{N}\leq
\frac{\sigma_n^{N-1}}{\sigma^{N}}.
\end{align}

Now, we focus on the case of a lossless cluster. By (\ref{Theorem6_01}), we notice that in this case, $\sigma_\mathrm{c}>0$ and thus, the interaction between the single-scatterer fields adds flux to the direct intensity. This fact comes in stark contrast with the behavior of the corresponding quantity (total interaction SCS) for multilayered scatterers, which can take negative values \cite{SAPM,QAM1,OJAP}. 

Furthermore, a straight-forward implication of (\ref{Theorem6_01}) and (\ref{OSCS}) is that
\begin{equation}\label{corol5_0}
    \mathrm{R}_{\mathrm{c}}=\frac{1}{1-\frac{\zeta_h}{\omega}\mathrm{Im}\left[\frac{A}{\beta_h}\overline{u_h^{\mathrm{sc}}}(\mathbf{a})\right]},
\end{equation}
and, thus, the ratio of the cluster-interaction cross section is related to the evaluation of the overall scattered field in the host medium at the location of the point source. Relation (\ref{corol5_0}) implies
\begin{corollary}
    The overall SCS is equal to the cluster-interaction cross section if and only if the field $\frac{A}{\beta_h}\overline{u^{\mathrm{sc}}}(\mathbf{a})$ is real. 
\end{corollary}

\section{Cluster of point-like scatterers}

\begin{figure}[!htb]
\centering
\includegraphics[width=0.65\textwidth]{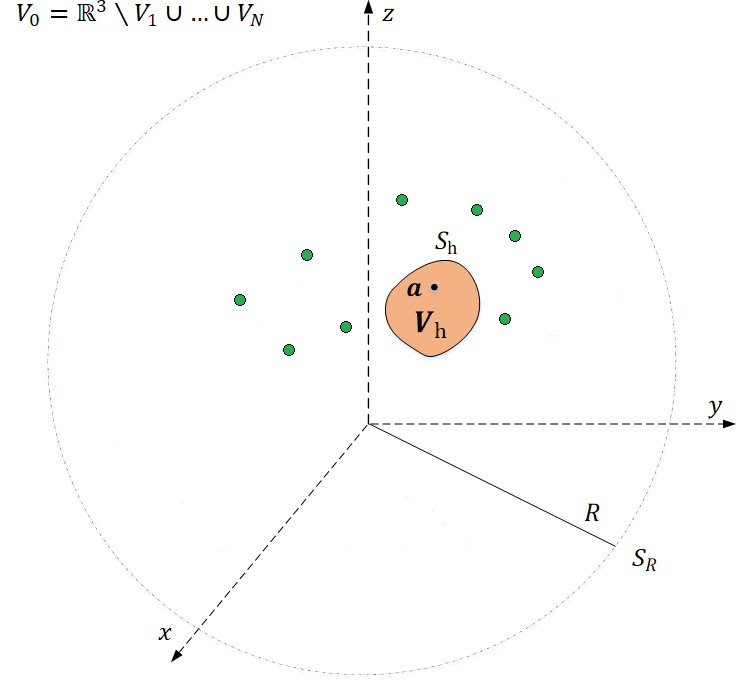}
\caption{A penetrable, host medium, $V_{h}$ with a point source in its interior radiated by a cluster of point-like scatterers.}
\label{fig:mult2}
\end{figure}

Now, we focus on the case when all the scatterers of the cluster are significantly smaller than the wavelength, with the exception of the host medium $V_h$; see Fig. \ref{fig:mult2}. Then, the overall field that impinges on $V_h$ constitutes an \emph{overall primary field} of the form \footnote{In this section, for convenience, we consider that the host medium $V_h$ is the $N$-th scatterer of the cluster.}
\begin{equation}
    \label{point-like_1}
    u_0^{\mathrm{pr}}(\mathbf{r})=\sum_{n=1}^{N-1}u_{0,n}^{\mathrm{pr}}(\mathbf{r})=\sum_{n=1}^{N-1}A_n\frac{e^{\mathrm{i}k_0\lvert\mathbf{r}-\mathbf{b}_n\rvert}}{\lvert\mathbf{r}-\mathbf{b}_n\rvert},
\end{equation}
where $\mathbf{b}_n$ is an internal point of $V_n$, and since $V_n$ is of arbitrary shape we can choose for convenience $\mathbf{b}_n$ to be the center of the circumscribing sphere surrounding $V_n$ \cite{Martin}. In this way, each scatterer of the cluster (excluding the host medium) is identified by the vector $\mathbf{b}_n$, so that these can be considered as ``point'' scatterers. We note that this sort of approximation is valid for sound-soft clusters, see \cite{Martin_MS, WM, JCOMP}.
Then, the multiple-scattering problem is actually that of a homogeneous penetrable medium excited by $N$ sources: $N-1$ lying in its exterior and one source in its interior. Thus, the total field in $V_0$ is $u^0=u_0^{\mathrm{pr}}+u_0^{\mathrm{sc}}$, while the total field in $V_h$ becomes $u^h=u_h^{\mathrm{pr}}+u_h^{\mathrm{sc}}$. Problems of this sort were treated for an arbitrary layered scatterer consisting of an arbitrary number of $P$ layers and an arbitrary number of $N$ sources in \cite{QAM1}; direct and inverse problems were analytically solved for a layered sphere in \cite{QAM2}. The properties of the energy transfer process for lossy scatterers were investigated in \cite{PRSA}. The effect of a collective scattered field induced by point scatterers impinging on a rigid body was investigated in \cite{MMS}. 

This formulation stems from Foldy's approximate method \cite{Foldy} for point-like scatterers, yet it has some mentionable differences. First, in Foldy's method, when the scatterers are approximated to be point-like, the multiple scattering problem is reduced to a free-space propagation problem. In our setting, the host-medium is not considered point-like and, thus, the formulation leads to a scattering problem by a single scatterer. The second difference is a consequence of the free-space propagation formulation. In particular, in Foldy's approximation method the only quantities radiating flux in the far-field are the primary fields. In our setting, there exists the field $u_0^{\mathrm{sc}}$ as well. Since we investigate the qualitative aspects of the energy transfer process, it suffices to consider the coefficients $A_n$ as ``strengths'' accounting for the power of the approximate single-scatterer fields impinging on the host medium. 
Still, without a priori assumptions about the nature of the point-like scatterrers, quantities $A_n$ cannot, in general, be observed \cite{Martin, Twersky2}.

With the above described formulation, the overall SCS is $\sigma=\sigma^{\mathrm{sc}}+\sigma^{\mathrm{pr}}$. The single-scatterer cross sections $\sigma_n^{\mathrm{pr}}$ are given by
\begin{equation}
\label{point-like_6}
    \sigma_n^{\mathrm{pr}}=\frac{1}{k_0^2}\int_{S^2}\lvert g_n^{\mathrm{pr}}(\hat{\mathbf{r}})\rvert^2\mathrm{d}s(\hat{\mathbf{r}}),
\end{equation}
which taking into account the far-field definition (\ref{far-field}) and relation (\ref{point-like_1}), becomes
\begin{equation}
\label{point-like_7}
    \sigma_n^{\mathrm{pr}}=4\pi\lvert A_n\rvert^2.
\end{equation}

The \emph{primary-interaction cross section} accounting for the flux induced by the interaction between the single-scatterer fields is defined as the sum of the interaction between different primary fields, i.e.,
\begin{equation}
    \label{point-like_8}
    \sigma_{\mathrm{R}}^{\mathrm{pr}}=\frac{1}{k_0^2}\sum_{\substack{k,m\\k\ne m}}\int_{S^2}g_k^{\mathrm{pr}}(\hat{\mathbf{r}})\overline{g_m^{\mathrm{pr}}}(\hat{\mathbf{r}})\mathrm{d}s(\hat{\mathbf{r}}).
\end{equation}

The following lemma provides a formula for the evaluation of the primary-interaction cross section.
\begin{lemma}
    The primary-interaction cross section $\sigma_{\mathrm{R}}^{\mathrm{pr}}$ is given by
    \begin{equation}
        \label{lemma1_0}
        \sigma_{\mathrm{R}}^{\mathrm{pr}}=4\pi\sum_{\substack{k,m\\k\ne m}}A_k\overline{A}_mj_0(k_0\lvert\mathbf{b}_k-\mathbf{b}_m\rvert),
    \end{equation}
with $j_0$ the spherical Bessel function of $0$-th order.
\end{lemma}
\textbf{Proof.}
Relation (\ref{point-like_8}), by means of (\ref{point-like_1}) and (\ref{far-field}), takes the form
\begin{equation}
    \label{point-like_9}
    \sigma_{\mathrm{R}}^{\mathrm{pr}}=\sum_{\substack{k,m\\k\ne m}}\int_{S^2}A_k\overline{A_m}e^{\mathrm{i}k_0\hat{\mathbf{r}}\cdot(\mathbf{b}_m-\mathbf{b}_k)}\mathrm{d}s(\hat{\mathbf{r}}).
\end{equation}
For the evaluation of the integral in (\ref{point-like_9}), we need only to evaluate one of its terms. By a change of variables, so that it holds 
$$z\cos{\theta}=\hat{\mathbf{r}}\cdot\frac{\mathbf{b}_m-\mathbf{b}_k}{\lvert\mathbf{b}_m-\mathbf{b}_k\rvert}$$
and taking into account property (10.54.2) of \cite{NIST}
\begin{equation}
    j_n(z)=\frac{(-\mathrm{i})^n}{2}\int_{0}^{\pi}e^{\mathrm{i}z\cos{\theta}}P_n(\cos{\theta})\sin{\theta}\mathrm{d}\theta,
\end{equation}
with $P_n$ the $n$-th Legendre polynomial, we obtain
\begin{equation}
    \label{point-like_10}
\int_{S^2}A_k\overline{A_m}e^{\mathrm{i}k_0\hat{\mathbf{r}}\cdot(\mathbf{b}_m-\mathbf{b}_k)}\mathrm{d}s(\hat{\mathbf{r}})=4\pi A_k\overline{A_m}j_0(k_0\lvert\mathbf{b}_m-\mathbf{b}_k\rvert).
\end{equation}
Equations (\ref{point-like_9}) and (\ref{point-like_10}) yield (\ref{lemma1_0}).
 \begin{flushright}$\Box$\end{flushright}

Thus, for the overall primary SCS combining (\ref{point-like_7}) and (\ref{lemma1_0}), we get
\begin{equation}
    \label{point-like_12}
    \sigma^{\mathrm{pr}}=\sum_{n=1}^{N-1}\sigma_n^{\mathrm{pr}}+\sigma_{\mathrm{R}}^{\mathrm{pr}}=4\pi\sum_{k=1}^{N-1}\sum_{m=1}^{N-1}A_k\overline{A}_mj_0(k_0\lvert\mathbf{b}_k-\mathbf{b}_m\rvert),
\end{equation} 
which recovers (53) of \cite{Martin}. In the absence of the host-medium, or if the host medium is also considered as a point-like scatterer, relation (\ref{point-like_12}) describes the overall SCS. In our setting, this is only a portion of the overall SCS. It accounts for the acoustic intensity (flux) radiated in the far-field, by the single-scatterer fields of the cluster. 
The secondary field $u_0^{\mathrm{sc}}$ in $V_0$  accounts for the intensity radiated in the far-field by the field elicited from the host-medium from both the presence of the point source in its interior and the excitation by the single-scatterer fields. Evidently, for a sparse cluster, it holds $\sigma_{\mathrm{R}}^{\mathrm{pr}}\simeq0$ and, thus, the primary SCS can be considered additive, while for a dense cluster, i.e., for $\lvert\mathbf{b}_k-\mathbf{b}_m\rvert<<1$ comprised by scatterers with $A_n\simeq A_0$, the following approximation holds 
$$\sigma^{\mathrm{pr}}\simeq4\pi\lvert A_0\rvert^2(N-1)^2.$$
%
%
\begin{theorem}
The overall SCS satisfies the relation
\begin{align}
\nonumber
\frac{\sigma}{\zeta_0}=\sum_{n=1}^{N-1}4\pi\lvert A_n\rvert^2\left( \frac{1}{\zeta_0}-\frac{1}{\omega\rho_0}\mathrm{Im}\left[\frac{u_0^{\mathrm{sc}}(\mathbf{b}_n)}{A_n}\right]\right)+\\\frac{4\pi\lvert A\rvert^2}{\zeta_h}\left(1-\mathrm{Im}\left[\frac{\zeta_h}{\beta_h}A\overline{u_h^{\mathrm{sc}}}(\mathbf{a})\right]\right)-2\omega\mathrm{Im}\left[\frac{\beta_h}{\rho_h}\right]\int_{V_h}\mathcal{K}^h(\mathbf{r})\mathrm{d}v(\mathbf{r}).
\label{theorem7_0}
\end{align}
\end{theorem}
\textbf{Proof.}
Let $B(\mathbf{b}_n,\epsilon)$ an open ball of radius $\epsilon$ centered at $\mathbf{b}_n$ surrounding scatterer $V_n$ and $S_R$ a sphere of radius $R$ surrounding the cluster. Denote $\Omega$ the subset of $\mathbb{R}^3$ with boundary $\partial\Omega=S_R-S_h-\cup_{n=1}^{N-1}\partial B(\mathbf{b}_n,\epsilon)$. Implementing Green's first scalar identity for $\mathbf{I}^0$ and $\overline{\mathbf{I}^0}$ in $\Omega$, we obtain
\begin{align}
\nonumber
\int_{S_R}\mathrm{Re}\left[\hat{\mathbf{r}}\cdot\mathbf{I}^0(\mathbf{r})\right]\mathrm{d}s(\mathbf{r})=\sum_{n=1}^{N-1}\int_{\partial B(\mathbf{b}_n,\epsilon)}\mathrm{Re}\left[\hat{\mathbf{r}}\cdot\mathbf{I}^0(\mathbf{r})\right]\mathrm{d}s(\mathbf{r})+\\\int_{S_h}\mathrm{Re}\left[\hat{\mathbf{r}}\cdot\mathbf{I}^0(\mathbf{r})\right]\mathrm{d}s(\mathbf{r})\label{theorem7_1}
\end{align}
For the acoustic intensity in $V_0$, (\ref{host_decomp_overall}) is valid. Letting $R\rightarrow\infty$ and utilizing (\ref{far-field}) and (\ref{eq:Sommerfeld}), we get
\begin{equation}
\label{theorem7_2}
\lim_{R\rightarrow\infty}\int_{S_R}\mathrm{Re}\left[\hat{\mathbf{r}}\cdot\mathbf{I}^0(\mathbf{r})\right]\mathrm{d}s(\mathbf{r})=\frac{\sigma}{\zeta_0}.
\end{equation}

On the other hand, for the integrals in the first term of the right-hand side of (\ref{theorem7_1}), it holds
\begin{align}
\nonumber
\int_{\partial B(\mathbf{b}_n,\epsilon)}\mathrm{Re}\left[\hat{\mathbf{r}}\cdot\mathbf{I}^0(\mathbf{r})\right]\mathrm{d}s(\mathbf{r})=\int_{\partial B(\mathbf{b}_n,\epsilon)}\mathrm{Re}\left[\hat{\mathbf{r}}\cdot\mathbf{I}_0^{\mathrm{pr}}(\mathbf{r})\right]\mathrm{d}s(\mathbf{r})+\\
\label{theorem7_3}\int_{\partial B(\mathbf{b}_n,\epsilon)}\mathrm{Re}\left[\hat{\mathbf{r}}\cdot\mathbf{I}_0^{\mathrm{sc}}(\mathbf{r})\right]\mathrm{d}s(\mathbf{r})+\int_{\partial B(\mathbf{b}_n,\epsilon)}\mathrm{Re}\left[\hat{\mathbf{r}}\cdot\mathbf{I}_0^{\mathrm{ext}}(\mathbf{r})\right]\mathrm{d}s(\mathbf{r}).
\end{align}
Since $u_0^{\mathrm{sc}}$ satisfies the Helmholtz equation in $V_0$, the second integral vanishes. For the integrands concerning the primary intensity $\mathbf{I}_0^{\mathrm{pr}}$, we have
\begin{align}
\nonumber
\mathrm{Re}\left[\hat{\mathbf{r}}\cdot\mathbf{I}_0^{\mathrm{pr}}(\mathbf{r})\right]=\frac{\mathrm{i}}{\omega\rho_0}\sum_{j=1}^{N-1}\sum_{k=1}^{N-1}\mathrm{Re}\left[u_j^{\mathrm{pr}}(\mathbf{r})\frac{\partial\overline{u_k^{\mathrm{pr}}}(\mathbf{r})}{\partial r}\right]=\\\nonumber\frac{1}{\omega\rho_0}\sum_{j=1}^{N-1}\sum_{k=1}^{N-1}A_j\overline{A_k}\Bigg(k_0\frac{\lvert\mathbf{r}-\mathbf{b}_j\rvert^2}{\lvert\mathbf{r}-\mathbf{b}_k\rvert^2}\cos{\left(k_0\left(\lvert\mathbf{r}-\mathbf{b}_j\rvert-\lvert\mathbf{r}-\mathbf{b}_k\rvert\right)\right)}+\\\frac{\lvert\mathbf{r}-\mathbf{b}_j\rvert}{\lvert\mathbf{r}-\mathbf{b}_k\rvert}\sin{\left(k_0\left(\lvert\mathbf{r}-\mathbf{b}_j\rvert-\lvert\mathbf{r}-\mathbf{b}_k\rvert\right)\right)}\Bigg).
\label{theorem7_31}
\end{align}
Applying the mean value theorem for each $B(\mathbf{b}_n,\epsilon)$, and letting $\epsilon\rightarrow0$, leads to
%
%
\begin{align}
\label{theorem7_6}
\lim_{\epsilon\rightarrow0}\int_{\partial B(\mathbf{b}_n,\epsilon)}\mathrm{Re}\left[\hat{\mathbf{r}}\cdot\mathbf{I}_0^{\mathrm{pr}}\right]\mathrm{d}s(\mathbf{r})=\frac{4\pi}{\zeta_0}\lvert A_n\rvert^2
\end{align}

For the integrands concerning the interaction intensity, it holds
\begin{align}
\label{theorem7_32}
&\mathrm{Re}\left[\hat{\mathbf{r}}\cdot\mathbf{I}_0^{\mathrm{ext}}(\mathbf{r})\right]=\frac{\mathrm{i}}{\omega\rho_0}\sum_{j=1}^{N-1}\sum_{k=1}^{N-1}\mathrm{Re}\left[u_j^{\mathrm{pr}}(\mathbf{r})\frac{\partial\overline{u_k^{\mathrm{sc}}}(\mathbf{r})}{\partial r}+u_j^{\mathrm{sc}}(\mathbf{r})\frac{\partial\overline{u_k^{\mathrm{pr}}}(\mathbf{r})}{\partial r}\right],
\end{align}
which after some algebra and applying the mean value theorem for each $B(\mathbf{b}_n,\epsilon)$ and $\epsilon\rightarrow0$, yields 
\begin{equation}
\label{theorem7_7}
\lim_{\epsilon\rightarrow0}\int_{\partial B(\mathbf{b}_n,\epsilon)}\mathrm{Re}\left[\hat{\mathbf{r}}\cdot\mathbf{I}_0^{\mathrm{ext}}\right]\mathrm{d}s(\mathbf{r})=-\frac{4\pi}{\omega\rho_0}\mathrm{Im}\left[\overline{A_n}u_0^{\mathrm{sc}}(\mathbf{b}_n)\right].
\end{equation}
Therefore, (\ref{theorem7_3}) takes the form
\begin{equation}
\label{theorem7_8}
\lim_{\epsilon\rightarrow0}\int_{\partial B(\mathbf{b}_n,\epsilon)}\mathrm{Re}\left[\hat{\mathbf{r}}\cdot\mathbf{I}^0\right]\mathrm{d}s(\mathbf{r})=\frac{4\pi}{\zeta_0}\lvert A_n\rvert^2-\frac{4\pi}{\omega\rho_0}\mathrm{Im}\left[\overline{A_n}u_0^{\mathrm{sc}}(\mathbf{b}_n)\right].
\end{equation}

Finally, for the second term in the right-hand side of (\ref{theorem7_1}), it holds
\begin{equation}
\label{theorem7_9}
\int_{S_h}\mathrm{Re}\left[\hat{\mathbf{r}}\cdot\mathbf{I}^0\right]\mathrm{d}s(\mathbf{r})=\int_{S_h}\mathrm{Re}\left[\hat{\mathbf{r}}\cdot\mathbf{I}^h\right]\mathrm{d}s(\mathbf{r}).
\end{equation}
Taking into account that the integral in the right-hand side of (\ref{theorem7_9}) is given by (\ref{Theorem6_10}), equation (\ref{theorem7_1}) by means of (\ref{theorem7_2}) and (\ref{theorem7_8}), yields (\ref{theorem7_0}). 
\begin{flushright}$\Box$\end{flushright}

\begin{remark}
Equation (\ref{theorem7_6}) implies that in the vicinity of each point scatterer, the induced flux results only from the scatterer's response, which under our formulation is interpreted as an individual primary intensity.
\end{remark}

\section{Conclusions}
The problem of the excitation of a cluster comprised by penetrable, lossy scatterers from a point-source enclosed in a penetrable, lossy host medium at close proximity with the rest of the cluster is formulated. The interaction intensity vectors, that quantify the sound flux induced by the interaction between fields elicited by different scatterers of the cluster and their corresponding cross sections, are defined and their role in multiple scattering problems involving lossy media is investigated. Scattering theorems that highlight the energy transfer process are established by means of Green's identities and techniques of asymptotic integration. These theorems connect the active interaction intensities with the cluster-interaction SCS and the reactive interaction intensities with the interaction Lagrangian density. The role of losses is also unveiled. Reductions for lossless scatterers are given. The specific contribution of the interaction and single-scatterer SCS is quantified by means of their ratios to the overall SCS and related physical bounds and limiting cases are derived. Finally, a formulation stemming from Foldy's method is proposed for the specific case of a cluster composed by point-like scatterers. The multiple scattering problem is reduced to a single-scattering problem and its theoretical implications are explicitly analyzed with an emphasis in the contribution of the single-scatterer to the overall scattered field.




\newpage

\bibliographystyle{elsarticle-num} 
\bibliography{main}

\begin{thebibliography}{10}
\expandafter\ifx\csname url\endcsname\relax
  \def\url#1{\texttt{#1}}\fi
\expandafter\ifx\csname urlprefix\endcsname\relax\def\urlprefix{URL }\fi
\expandafter\ifx\csname href\endcsname\relax
  \def\href#1#2{#2} \def\path#1{#1}\fi

\bibitem{Jacobsen-1991}
F.~Jacobsen, A note on instantaneous and time-averaged active and reactive sound intensity, J. Sound Vib. 147~(3) (1991) 489–496.
\newblock \href {https://doi.org/10.1016/0022-460X(91)90496-7} {\path{doi:10.1016/0022-460X(91)90496-7}}.

\bibitem{JASA2}
D.~Stanzial, G.~Sacchi, G.~Schiffrer, On the physical meaning of the power factor in acoustics, J. Acoust. Soc. Am. 131~(1) (2012) 269–280.
\newblock \href {https://doi.org/10.1121/1.3664051} {\path{doi:10.1121/1.3664051}}.

\bibitem{Stanzial}
D.~Stanzial, N.~Prodi, G.~Schiffrer, Reactive acoustic intensity for energy fields and energy polarization, J. Acoust. Soc. Am. 99~(4) (1996) 1868–1876.
\newblock \href {https://doi.org/10.1121/1.415369} {\path{doi:10.1121/1.415369}}.

\bibitem{Nolan}
M.~Nolan, J.~Davy, Two definitions of the inner product of modes and their use in calculating non-diffuse reverberant sound fields, J. Acoust. Soc. Am. 145~(6) (2019) 3330--3340.
\newblock \href {https://doi.org/10.1121/1.5109662} {\path{doi:10.1121/1.5109662}}.

\bibitem{JASA}
W.~Duan, R.~Kirby, J.~Prisutova, K.~Horoshenkov, Measurement of complex acoustic intensity in an acoustic waveguide, J. Acoust. Soc. Am. 134~(5) (2013) 3674–3685.
\newblock \href {https://doi.org/10.1121/1.4821214} {\path{doi:10.1121/1.4821214}}.

\bibitem{Zhang}
G.~Zhang, C.~Zheng, W.~Lin, Steering acoustic intensity estimator using a single acoustic vector hydrophone, Journal of Sensors 2018 (2018) 8526092.
\newblock \href {https://doi.org/10.1155/2018/8526092} {\path{doi:10.1155/2018/8526092}}.

\bibitem{Aramini}
R.~Aramini, G.~Caviglia, G.~Giorgi, The role of point sources and their power fluxes in the linear sampling method, SIAM J. Appl. Math. 71~(4) (2011) 1044–1069.
\newblock \href {https://doi.org/10.1137/100814780} {\path{doi:10.1137/100814780}}.

\bibitem{Chardon}
G.~Chardon, Theoretical analysis of beamforming steering vector formulations for acoustic source localization, J. Sound Vib. 517 (2022) 116544.
\newblock \href {https://doi.org/10.1016/j.jsv.2021.116544} {\path{doi:10.1016/j.jsv.2021.116544}}.

\bibitem{Jacobsen}
F.~Jacobsen, A.~Molares, Ensemble statistics of active and reactive sound intensity in reverberation rooms, J. Acoust. Soc. Am. 129~(1) (2011) 211–218.
\newblock \href {https://doi.org/10.1121/1.3514425} {\path{doi:10.1121/1.3514425}}.

\bibitem{Meissner}
M.~Meissner, Numerical investigation of acoustic field in enclosures: Evaluation of active and reactive components of sound intensity, J. Sound Vib. 338 (2015) 154--168.
\newblock \href {https://doi.org/10.1016/j.jsv.2021.116544} {\path{doi:10.1016/j.jsv.2021.116544}}.

\bibitem{PRSA}
A.~Kalogeropoulos, N.~L. Tsitsas, Analysis of the energy transfer process for multiple scattering problems involving lossy media, Proc. R. Soc. A 479 (2023) 20230513.
\newblock \href {https://doi.org/10.1098/rspa.2023.0513} {\path{doi:10.1098/rspa.2023.0513}}.

\bibitem{Twersky}
V.~Twersky, Multiple scattering by arbitrary configurations in three dimensions, J. Math. Phys. 3~(1) (1962) 83–91.
\newblock \href {https://doi.org/10.1063/1.1703791} {\path{doi:10.1063/1.1703791}}.

\bibitem{Spyropoulos}
C.~Athanasiadis, P.~A. Martin, A.~Spyropoulos, I.~G. Stratis, Scattering relations for point sources: Acoustic and electromagnetic waves, Journal of Mathematical Physics 43 (2002) 5683--5697.
\newblock \href {https://doi.org/10.1063/1.1509089} {\path{doi:10.1063/1.1509089}}.

\bibitem{Tsitsas-SAPM}
C.~Athanasiadis, N.~L. Tsitsas, Scattering theorems for acoustic excitation of a layered obstacle by an interior point source, Studies in Applied Mathematics 118 (2007) 397--418.
\newblock \href {https://doi.org/10.1111/j.1365-2966.2007.00375.x} {\path{doi:10.1111/j.1365-2966.2007.00375.x}}.

\bibitem{SAPM}
A.~Kalogeropoulos, N.~L. Tsitsas, Electromagnetic interactions of dipole distributions with a stratified medium: Power fluxes and scattering cross sections, Stud. Appl. Math. 148~(3) (2022) 1040--1068.
\newblock \href {https://doi.org/10.1111/sapm.12469} {\path{doi:10.1111/sapm.12469}}.

\bibitem{Marston}
P.~Marston, Generalized optical theorem for scatterers having inversion symmetry: Applications to acoustic backscattering, J. Acoust. Soc. Am. 109~(4) (2001) 1291--1295.
\newblock \href {https://doi.org/10.1121/1.1352082} {\path{doi:10.1121/1.1352082}}.

\bibitem{Wu}
J.~Wu, A.~Liu, H.~Chen, T.~Chen, Multiple scattering of a spherical acoustic wave from fluid spheres, J. Sound Vib. 290~(1-2) (2006) 17--33.
\newblock \href {https://doi.org/10.1016/j.jsv.2005.03.015} {\path{doi:10.1016/j.jsv.2005.03.015}}.

\bibitem{RSL}
A.~Kalogeropoulos, N.~L. Tsitsas, On interactions between spherical waves in multiple scattering by an all-dielectric cluster, URSI Radio Science Letters 3 (2021).
\newblock \href {https://doi.org/10.46620/21-0034} {\path{doi:10.46620/21-0034}}.

\bibitem{Martin}
P.~A. Martin, Multiple scattering and scattering cross sections, J. Acoust. Soc. Am. 143~(2) (2019) 995--1002.
\newblock \href {https://doi.org/10.1121/1.5024361} {\path{doi:10.1121/1.5024361}}.

\bibitem{Martin2}
P.~A. Martin, Quadratic quantities in acoustics: scattering cross section and radiation force, Wave Motion 86 (2019) 63--78.
\newblock \href {https://doi.org/10.1016/j.wavemoti.2018.12.009} {\path{doi:10.1016/j.wavemoti.2018.12.009}}.

\bibitem{QAM1}
A.~Kalogeropoulos, N.~L. Tsitsas, Excitation of a layered medium by {$N$} sources: Scattering relations, interaction cross sections and physical bounds, Quart. Appl. Math. 79~(2) (2021) 335--356.
\newblock \href {https://doi.org/10.1090/qam/1581} {\path{doi:10.1090/qam/1581}}.

\bibitem{LFS}
G.~Dassios, R.~Kleinman, Low Frequency Scattering, Clarendon Press, Oxford, 2000.

\bibitem{Martin_MS}
P.~A. Martin, Multiple Scattering: Interaction of time-harmonic waves with N obstacles, Cambridge University Press, 2006.

\bibitem{Colton-Kress-IEM}
D.~Colton, R.~Kress, Integral {E}quation {M}ethods in {S}cattering {T}heory, SIAM, 2013.

\bibitem{Rumsey}
V.~H. Rumsey, Reaction concept in electromagnetic theory, Phys. Rev. 94~(6) (1954) 1483--1491.
\newblock \href {https://doi.org/10.1103/PhysRev.94.1483} {\path{doi:10.1103/PhysRev.94.1483}}.

\bibitem{Sihvola}
I.~Lindell, A.~Sihvola, Rumsey’s reaction concept generalized, Prog. Electromagn. Res. Lett. 89 (2020) 1--6.
\newblock \href {https://doi.org/10.2528/PIERL19091705} {\path{doi:10.2528/PIERL19091705}}.

\bibitem{OJAP}
A.~Kalogeropoulos, N.~L. Tsitsas, Analysis of interaction scattering cross sections and their physical bounds for multiple-dipole stimulation of a three-dimensional layered medium, IEEE Open. J. Antennas Propag. 2 (2021) 506--520.
\newblock \href {https://doi.org/10.1109/OJAP.2021.3070199} {\path{doi:10.1109/OJAP.2021.3070199}}.

\bibitem{WM}
M.~Cassier, C.~Hazard, Multiple scattering of acoustic waves by small sound-soft obstacles in two dimensions: Mathematical justification of the foldy–lax model, Wave Motion 50 (2013) 18--28.
\newblock \href {https://doi.org/10.1016/j.wavemoti.2012.06.001} {\path{doi:10.1016/j.wavemoti.2012.06.001}}.

\bibitem{JCOMP}
K.~Huang, P.~Li, H.~Zhao, An efficient algorithm for the generalized foldy–lax formulation, J. Comput. Phys. 234 (2013) 376--398.
\newblock \href {https://doi.org/10.1016/j.jcp.2012.09.027} {\path{doi:10.1016/j.jcp.2012.09.027}}.

\bibitem{QAM2}
A.~Kalogeropoulos, N.~L. Tsitsas, Excitation of a layered sphere by {$N$} acoustic sources: Exact solutions, low-frequency approximations and inverse problems, Quart. Appl. Math. 81~(1) (2023) 141--173.
\newblock \href {https://doi.org/10.1090/qam/1632} {\path{doi:10.1090/qam/1632}}.

\bibitem{MMS}
G.~Hu, A.~Mantile, M.~Sini, Direct and inverse acoustic scattering by a collection of extended and point-like scatterers, Multiscale Model. Sim. 12~(3) (2014) 996--1027.
\newblock \href {https://doi.org/10.1137/130932107} {\path{doi:10.1137/130932107}}.

\bibitem{Foldy}
L.~L. Foldy, The multiple scattering of waves. i. general theory of isotropic scattering by randomly distributed scatterers, Phys. Rev. 67 (1945) 107--119.
\newblock \href {https://doi.org/10.1103/PhysRev.67.107} {\path{doi:10.1103/PhysRev.67.107}}.

\bibitem{Twersky2}
V.~Twersky, Multiple scattering by finite regular arrays of resonators, J. Acoust. Soc. Am. 87 (1990) 25--41.
\newblock \href {https://doi.org/10.1121/1.399292} {\path{doi:10.1121/1.399292}}.

\bibitem{NIST}
Digital library of mathematical functions, available at dlmf.nist.gov, (last viewed 10/10/2024).

\end{thebibliography}






\end{document}